\documentclass[12pt,a4paper]{article}

\usepackage{latexsym}
\usepackage{amssymb,exscale}
\usepackage[centertags]{amsmath}

\renewcommand{\(}{\bigl(}
\renewcommand{\)}{\bigr)\vphantom{)}}
\newcommand{\ip}[2]{\,\langle\,#1\,|\,#2\,\rangle\,}
\newcommand{\ket}[1]{\,|#1\rangle\,}
\renewcommand{\Re}{\operatorname{Re}}
\renewcommand{\Im}{\operatorname{Im}}
\newcommand{\cT}{\mathcal T}
\newcommand{\cH}{\mathcal H}
\newcommand{\xmin}{{x_{\text{min}}}}
\def\emailwww#1#2{\par\qquad {\tt #1}\par\qquad {\tt #2}\medskip}

\begin{document}

\title{The quantum algorithm of Kieu\\
  does not solve the Hilbert's tenth problem}
\author{Boris Tsirelson}
\date{}

\maketitle

\begin{abstract}
Recently T.~Kieu \cite{Ki} claimed a quantum algorithm computing some
functions beyond the Church-Turing class. He notes that ``it is in
fact widely believed that quantum computation cannot offer anything
new about computability'' and claims the opposite. However, his
quantum algorithm does not work, which is the point of my short
note. I still believe that quantum computation leads to new complexity
but retains the old computability.
\end{abstract}

\section{The algorithm of Kieu}

Obtaining the ground state of a Hamiltonian could be treated as a
basic operation (feasible by definition), but such an algorithm should
be called algorithm with a ground-state oracle rather than quantum
algorithm. By the quantum algorithm of Kieu I mean the method
described in Sect.\ ``Adiabatic evolution'' of \cite{Ki}.\footnote{%
 Though, he did not specify the initial Hamiltonian $ H_I $. We'll see
 that his algorithm fails for every $ H_I $.}

Here is the relevant portion of the algorithm. In order to find the
ground state $ \ket g $ of a given Hamiltonian $ H_P $, one uses the
adiabatic evolution from the known ground state $ \ket{g_I} $ of an
initial Hamiltonian $ H_I $:
\[
\ket g = \lim_{T\to\infty} \cT \exp \bigg( -i \int_0^T \cH(t) \, dt
\bigg) \ket{g_I}
\]
up to a phase; that is, a state vector $ \ket{g(t)} $ evolves by the
Schr\"odinger equation
\begin{gather*}
\frac{d}{dt} \ket{g(t)} = -i \cH(t) \ket{g(t)} \quad \text{for } 0 \le
 t \le T \, , \\
\ket{g(0)} = \ket{g_I},
\end{gather*}
where
\begin{gather*}
\cH(t) = \bigg( 1 - \frac t T \bigg) H_I + \frac t T H_P \quad
 \text{for } 0 \le t \le T \, , \\
\cH(0) = H_I \, , \qquad \cH(T) = H_P \, ,
\end{gather*}
giving $ \ket{g(T)} \approx \ket g $ (up to a phase) provided that $ T
$ is large enough.

\section{The goal of the algorithm}

The quantum algorithm of Kieu is intended for finding the global
minimum of a function\footnote{%
 Kieu considers only polynomials $ P $, which is essential when
 implementing the Hamiltonian $ H_P $, but irrelevant when finding the
 minimum by adiabatic evolution.}
$ P(x,y,z) $ of, say, three variables $ x,y,z $ each running over
nonnegative integers $ \{ 0,1,2,\dots \} $.\footnote{%
 For example, the function $ P(x,y,z) = \( (x+1)^3 + (y+1)^3 - (z+1)^3
 \)^2 $ is evidently related to the Fermat's last theorem.}
To this end, the adiabatic evolution finds the ground state of the
Hamiltonian $ H_P $ whose eigenvalues are the numbers $ P(x,y,z) $.

Let us use a single-variable function $ P(x) $, just for simplifying
notation. Our Hilbert space is spanned by basis vectors $ \ket x $ for
$ x \in \{ 0,1,2,\dots \} $, and $ H_P $ is a diagonal operator,
\[
H_P \ket x = P(x) \ket x \quad \text{for } x \in \{ 0,1,2,\dots \} \, .
\]

The user of the algorithm has no upper bound for the minimizer $ \xmin
$ of the function $ P $. You see, if an upper bound is known, then the
global minimum can be found classically, just by computing and
comparing a finite number of values $ P(x) $. The goal of the
algorithm is finding the global minimum
\[
P(\xmin) = \min_{x\in\{0,1,2,\dots\}} P(x)
\]
and the minimizer $ \xmin $, during a time $ T $ that does not depend
on $ P $. That is crucial: \emph{the whole infinite domain $ \{
0,1,2,\dots \} $ must be effectively examined during the finite time $
T $.} According to Kieu, that is possible due to infinite dimension of
the Hilbert space.

If $ H_I $ is also diagonal in the given basis $ \( \ket x
\)_{x=0,1,\dots} $, then clearly $ \ket g = \ket{g_I} $ (up to a
phase); nothing happens. This is why operators $ H_I $ and $ H_P $
must be non-commutative, which is noted by Kieu, Sect.\ ``Gap
estimation'' of \cite{Ki}. Matrix elements of $ H_I $ must connect
basis vectors. However, they cannot effectively connect first basis
vectors with remote basis vectors $ \ket x $ for arbitrarily large $ x
$, as one could guess. That is the obstacle. See the next section for
a proof.

\section{The goal is not achieved}

Consider such a function $ P(\cdot) $:\footnote{%
 Kieu considers only nonnegative functions. Well, my argument works
 also for the nonnegative function $ P(x)+1 $.}
\[
P(x) = \begin{cases}
 -1 &\text{if $ x = \xmin $},\\
 0 & \text{otherwise}.
\end{cases}
\]
In order to disclose its minimum at $ \xmin $ (unknown to the user),
the adiabatic evolution is harnessed:
\begin{gather*}
\frac{d}{dt} \ket{g(t)} = -i \cH(t) \ket{g(t)} \quad \text{for } 0 \le
 t \le T \, , \\
\ket{g(0)} = \ket{g_I}, \\
\cH(t) = \bigg( 1 - \frac t T \bigg) H_I + \frac t T H_P \, .
\end{gather*}
Compare $ \ket{g(\cdot)} $ with the solution $ \ket{g_0(\cdot)} $ of a
simpler equation
\begin{gather*}
\frac{d}{dt} \ket{g_0(t)} = -i \cH_0(t) \ket{g_0(t)}, \quad
 \ket{g_0(0)} = \ket{g_I}, \\
\cH_0(t) = \bigg( 1 - \frac t T \bigg) H_I \, .
\end{gather*}
Note that $ \ket{g_0(t)} = \ket{g_I} $ up to a phase (varying in
time), since $ \ket{g_I} $ is an eigenvector for $ H_I $, therefore
for $ \cH_0(t) $. We have
\begin{align*}
& \frac{d}{dt} \ket{ g(t) - g_0(t) } = -i \cH(t) \ket{ g(t) - g_0(t) }
 + i \frac t T H_P \ket{g_0(t)}; \\
& \frac{d}{dt} \| \ket{ g(t) - g_0(t) } \|^2 = 2 \Re \ip{ g'(t) -
 g'_0(t) }{ g(t) - g_0(t) } = \\
& \quad = -2 \Im \ip{ \tfrac t T H_P g_0(t) }{ g(t) - g_0(t) } \le 2
 \| H_P \ket{ g_0(t) } \| \cdot \| \ket{ g(t) - g_0(t) } \| \, ; \\
& \frac{d}{dt} \| \ket{ g(t) - g_0(t) } \| \le \| H_P \ket{g_0(t)} \|
 = \| H_P \ket{g_I} \| \, ; \\
& \| \ket{g(t)} - \ket{g_0(t)} \| \le T \cdot \| H_P g_I \| \, .
\end{align*}
The arbitrary parameter $ \xmin $, unknown to the user, does not
influence $ T $, $ H_I $ and $ g_I $. However,
\[
\| H_P \ket{g_I} \| = \big| \ip{ \xmin }{ g_I } \big|
\xrightarrow[\xmin\to\infty]{} 0 \, .
\]
For large $ \xmin $ we see that $ \ket{ g(T) } $ is close to $ \ket{
g_0(T) } $; the latter, being equal to $ \ket{g_I} $ up to a phase,
cannot be close to $ \ket{\xmin} $. The adiabatic evolution fails to
disclose the minimum.

\bigskip
\filbreak
{
\small
\begin{sc}
\parindent=0pt\baselineskip=12pt

School of Mathematics, Tel Aviv Univ., Tel Aviv
69978, Israel
\smallskip
\emailwww{tsirel@tau.ac.il}
{http://www.math.tau.ac.il/$\sim$tsirel/}
\end{sc}
}
\filbreak

\end{document}